% PLAIN TEX!

% Look out for 8th bit; if you can't handle it, download ps instead.

% MACROS (reduced version for hep-th)

%
% TEXT
%

% fonts

\def\famname{
 \textfont0=\textrm \scriptfont0=\scriptrm
 \scriptscriptfont0=\sscriptrm
 \textfont1=\textmi \scriptfont1=\scriptmi
 \scriptscriptfont1=\sscriptmi
 \textfont2=\textsy \scriptfont2=\scriptsy \scriptscriptfont2=\sscriptsy
 \textfont3=\textex \scriptfont3=\textex \scriptscriptfont3=\textex
 \textfont4=\textbf \scriptfont4=\scriptbf \scriptscriptfont4=\sscriptbf
 \skewchar\textmi='177 \skewchar\scriptmi='177
 \skewchar\sscriptmi='177
 \skewchar\textsy='60 \skewchar\scriptsy='60
 \skewchar\sscriptsy='60
 \def\rm{\fam0 \textrm} \def\bf{\fam4 \textbf}}
\def\sca#1{scaled\magstep#1} \def\scah{scaled\magstephalf} 
\def\twelvepoint{
 \font\textrm=cmr12 \font\scriptrm=cmr8 \font\sscriptrm=cmr6
 \font\textmi=cmmi12 \font\scriptmi=cmmi8 \font\sscriptmi=cmmi6 
 \font\textsy=cmsy10 \sca1 \font\scriptsy=cmsy8
 \font\sscriptsy=cmsy6
 \font\textex=cmex10 \sca1
 \font\textbf=cmbx12 \font\scriptbf=cmbx8 \font\sscriptbf=cmbx6
 \font\it=cmti12
 \font\sectfont=cmbx12 \sca1
 \font\refrm=cmr10 \scah \font\refit=cmti10 \scah
 \font\refbf=cmbx10 \scah
 \def\twelverm{\textrm} \def\twelveit{\it} \def\twelvebf{\textbf}
 \famname \textrm 
 \voffset=.04in \hoffset=.21in
 \normalbaselineskip=18pt plus 1pt \baselineskip=\normalbaselineskip
 \parindent=21pt
 \setbox\strutbox=\hbox{\vrule height10.5pt depth4pt width0pt}}

% primes and other "@"garbage

\catcode`@=11

{\catcode`\'=\active \def'{{}^\bgroup\prim@s}}

\def\screwcount{\alloc@0\count\countdef\insc@unt}   % for stupid
\def\screwdimen{\alloc@1\dimen\dimendef\insc@unt} % \outer errors
\def\screwbox{\alloc@4\box\chardef\insc@unt}

\catcode`@=12

% Text style parameters ("textile"?)

\overfullrule=0pt			% gets rid of stupid black boxes
\voffset=.04in \hoffset=.21in
\vsize=9in \hsize=6in
\parskip=\medskipamount	% space between paragraphs
\lineskip=0pt				% minimum box separation
\normalbaselineskip=18pt plus 1pt \baselineskip=\normalbaselineskip
\abovedisplayskip=1.2em plus.3em minus.9em % space above equation
\belowdisplayskip=1.2em plus.3em minus.9em	% " below
\abovedisplayshortskip=0em plus.3em	% " above when no overlap
\belowdisplayshortskip=.7em plus.3em minus.4em	% " below
\parindent=21pt
\setbox\strutbox=\hbox{\vrule height10.5pt depth4pt width0pt}
\def\makefootline{\baselineskip=30pt \line{\the\footline}}
\footline={\ifnum\count0=1 \hfil \else\hss\twelverm\folio\hss \fi}
\pageno=1

% Boxes drawn around phrases or paragraphs

\def\boxit#1{\leavevmode\thinspace\hbox{\vrule\vtop{\vbox{
	\hrule\kern1pt\hbox{\vphantom{\bf/}\thinspace{\bf#1}\thinspace}}
	\kern1pt\hrule}\vrule}\thinspace}
\def\Boxit#1{\noindent\vbox{\hrule\hbox{\vrule\kern3pt\vbox{
	\advance\hsize-7pt\vskip-\parskip\kern3pt\bf#1
	\hbox{\vrule height0pt depth\dp\strutbox width0pt}
	\kern3pt}\kern3pt\vrule}\hrule}}

% Box at relative coordinates (x,y)

\def\put(#1,#2)#3{\screwdimen\unit  \unit=1in
	\vbox to0pt{\kern-#2\unit\hbox{\kern#1\unit
	\vbox{#3}}\vss}\nointerlineskip}

% PAPER:
%	\paper
%
%	"title"
%
%	"authors"
%
%	"preprint number"
%
%	"date"
%
%	"abstract"
%
%	"text"
%
%	\bye

\def\\{\hfil\break}

\def\center{\leftskip=0pt plus 1fill \rightskip=\leftskip \parindent=0pt
 \def\textindent##1{\par\hangindent21pt\footrm\noindent\hskip21pt
 \llap{##1\enspace}\ignorespaces}\par}
 % use as {\center ... \par}, shorten lines with \\
\def\unnarrower{\leftskip=0pt \rightskip=\leftskip}
\def\thetitle#1#2#3#4#5{
 \font\titlefont=cmbx12 \sca2 \font\footrm=cmr10 \font\footit=cmti10
  \twelverm
	{\hbox to\hsize{#4 \hfill ITP-SB-#3}}\par
	\vskip.8in minus.1in {\center\baselineskip=1.44\normalbaselineskip
 {\titlefont #1}\par}{\center\baselineskip=\normalbaselineskip
 \vskip.5in minus.2in #2
	\vskip1.4in minus1.2in {\twelvebf ABSTRACT}\par}
 \vskip.1in\par
 \narrower\par#5\par\unnarrower\vskip3.5in minus2.3in\eject}
\def\paper\par#1\par#2\par#3\par#4\par#5\par{\twelvepoint
	\thetitle{#1}{#2}{#3}{#4}{#5}} 
\def\author#1#2{#1 \vskip.1in {\twelveit #2}\vskip.1in}
\def\ITP{Institute for Theoretical Physics\\
	State University of New York, Stony Brook, NY 11794-3840}

% Page and section headings and reference stuff (also, \refs below)

\def\sect#1\par{\par\ifdim\lastskip<\medskipamount
	\bigskip\medskip\goodbreak\else\nobreak\fi
	\noindent{\sectfont{#1}}\par\nobreak\medskip} % see Ü below 
\def\itemize#1 {\item{[#1]}}	% see £ below; also use \itemitem 
\def\vol#1 {{\refbf#1} }		 % see É below

\def\ref#1{\setbox0=\hbox{M}$\vbox to\ht0{}^{#1}$}

% Journal abbreviations

\def\NP #1 {{\refit Nucl. Phys.} {\refbf B{#1}} }
\def\PL #1 {{\refit Phys. Lett.} {\refbf{#1}} }
\def\PR #1 {{\refit Phys. Rev. Lett.} {\refbf{#1}} }
\def\PRD #1 {{\refit Phys. Rev.} {\refbf D{#1}} }

% More nitpicking

\hyphenation{pre-print}
\hyphenation{quan-ti-za-tion}

%
% MATH (mostly)
%

% accent over:
\def\on#1#2{{\buildrel{\mkern2.5mu#1\mkern-2.5mu}\over{#2}}}
\def\dt#1{\on{\hbox{\bf .}}{#1}}                % (big) dot over: seeÀ   below  
\def\ddt#1{\on{\hbox{\bf .\kern-1pt.}}#1}    % À À   (see below)
\def\slap#1#2{\setbox0=\hbox{$#1{#2}$}
	#2\kern-\wd0{\hbox to\wd0{\hfil$#1{/}$\hfil}}}
\def\sla#1{\mathpalette\slap{#1}}                % slash: see Ö   below
\def\bop#1{\setbox0=\hbox{$#1M$}\mkern1.5mu
	\vbox{\hrule height0pt depth.04\ht0
	\hbox{\vrule width.04\ht0 height.9\ht0 \kern.9\ht0
	\vrule width.04\ht0}\hrule height.04\ht0}\mkern1.5mu}
\def\bo{{\mathpalette\bop{}}}                        % box: see õ below
\def~{\widetilde} % tilde key; use Option-N for accent in text & math,
	% Option-' for sim, Option-0 for math space (¼)
\mathcode`\*="702A                  % * now always complex conjugate
\def\in{\relax\ifmmode\mathchar"3232\else{\refit in\/}\fi} % ã below 
\def\f#1#2{{\textstyle{#1\over#2}}}	   % fraction
\def\half{{\textstyle{1\over{\raise.1ex\hbox{$\scriptstyle{2}$}}}}}

\catcode`\^^?=13				    % Clear
\catcode128=13 \def €{\"A}                 % Option-u A
\catcode129=13 \def {\AA}                 % Option-A
\catcode130=13 \def '{\c}           	   % Option-C (cedilla)
\catcode131=13 \def ƒ{\'E}                   % Option-e E
\catcode132=13 \def "{\~N}                   % Option-n N
\catcode133=13 \def …{\"O}                 % Option-u O
\catcode134=13 \def †{\"U}                  % Option-u U
\catcode135=13 \def ‡{\'a}                  % Option-e a
\catcode136=13 \def ˆ{\`a}                   % Option-`  a
\catcode137=13 \def ‰{\^a}                 % Option-i a
\catcode138=13 \def Š{\"a}                 % Option-u a
\catcode139=13 \def ‹{\~a}                   % Option-n a
\catcode140=13 \def Œ{\alpha}            % Option-a
\catcode141=13 \def {\chi}                % Option-c
\catcode142=13 \def Ž{\'e}                   % Option-e e
\catcode143=13 \def {\`e}                    % Option-`  e
\catcode144=13 \def {\^e}                  % Option-i e
\catcode145=13 \def '{\"e}                % Option-u e
\catcode146=13 \def '{\'\i}                 % Option-e i
\catcode147=13 \def "{\`\i}                  % Option-`  i
\catcode148=13 \def "{\^\i}                % Option-i i
\catcode149=13 \def •{\"\i}                % Option-u i
\catcode150=13 \def –{\~n}                  % Option-n n
\catcode151=13 \def —{\'o}                 % Option-e o
\catcode152=13 \def ˜{\`o}                  % Option-`  o
\catcode153=13 \def ™{\^o}                % Option-i o
\catcode154=13 \def š{\"o}                 % Option-u o
\catcode155=13 \def ›{\~o}                  % Option-n o
\catcode156=13 \def œ{\'u}                  % Option-e u
\catcode157=13 \def {\`u}                  % Option-`  u
\catcode158=13 \def ž{\^u}                % Option-i u
\catcode159=13 \def Ÿ{\"u}                % Option-u u
\catcode160=13 \def  {\tau}               % Option-t
\catcode161=13 \mathchardef ¡="2203     % Option-* (TeX's usual eq. *)
\catcode162=13 \def ¢{\oplus}           % Option-4
\catcode163=13 \def £{\relax\ifmmode\to\else\itemize\fi} % Option-3
\catcode164=13 \def ¤{\subset}	  % Option-6
\catcode165=13 \def ¥{\infty}           % Option-8
\catcode166=13 \def ¦{\mp}                % Option-7
\catcode167=13 \def §{\sigma}           % Option-s
\catcode168=13 \def ¨{\rho}               % Option-r
\catcode169=13 \def ©{\gamma}         % Option-g
\catcode170=13 \def ª{\leftrightarrow} % Option-2 ; Option-E (acute) :
\catcode171=13 \def «{\relax\ifmmode\acute\else\expandafter\'\fi}
\catcode172=13 \def ¬{\relax\ifmmode\expandafter\ddt\else\expandafter\"\fi}
\catcode173=13 \def ­{\equiv}            % Option-= ; ^ Option-U (umlaudt)
\catcode174=13 \def ®{\approx}          % Option-"
\catcode175=13 \def ¯{\Omega}          % Option-O
\catcode176=13 \def °{\otimes}          % Option-5
\catcode177=13 \def ±{\ne}                 % Option-+
\catcode178=13 \def ²{\le}                   % Option-,
\catcode179=13 \def ³{\ge}                  % Option-.
\catcode180=13 \def ´{\upsilon}          % Option-y
\catcode181=13 \def µ{\mu}                % Option-m
\catcode182=13 \def ¶{\delta}             % Option-d
\catcode183=13 \def ·{\epsilon}          % Option-w
\catcode184=13 \def ¸{\Pi}                  % Option-P
\catcode185=13 \def ¹{\pi}                  % Option-p
\catcode186=13 \def º{\beta}               % Option-b
\catcode187=13 \def »{\partial}           % Option-9
\catcode188=13 \def ¼{\nobreak\ }       % Option-0
\catcode189=13 \def ½{\zeta}               % Option-z
\catcode190=13 \def ¾{\sim}                 % Option-'
\catcode191=13 \def ¿{\omega}           % Option-o
\catcode192=13 \def À{\dt}                     % Option-?
\catcode193=13 \def Á{\gets}                % Option-1
\catcode194=13 \def Â{\lambda}           % Option-l
\catcode195=13 \def Ã{\nu}                   % Option-v
\catcode196=13 \def Ä{\phi}                  % Option-f
\catcode197=13 \def Å{\xi}                     % Option-x
\catcode198=13 \def Æ{\psi}                  % Option-j
\catcode199=13 \def Ç{\int}                    % Option-\
\catcode200=13 \def È{\oint}                 % Option-|
\catcode201=13 \def É{\relax\ifmmode\cdot\else\vol\fi}    % Option-;
\catcode202=13 \def Ê{\relax\ifmmode\,\else\thinspace\fi}
\catcode203=13 \def Ë{\`A}                      % Option-`  A ; ^ Option-space
\catcode204=13 \def Ì{\~A}                      % Option-n A
\catcode205=13 \def Í{\~O}                      % Option-n O
\catcode206=13 \def Î{\Theta}              % Option-Q
\catcode207=13 \def Ï{\theta}               % Option-q; Option-- :
\catcode208=13 \def Ð{\relax\ifmmode\bar\else\expandafter\=\fi}
\catcode209=13 \def Ñ{\overline}             % Option-_
\catcode210=13 \def Ò{\langle}               % Option-[
\catcode211=13 \def Ó{\relax\ifmmode\{\else\ital\fi}      % Option-{
\catcode212=13 \def Ô{\rangle}               % Option-]
\catcode213=13 \def Õ{\}}                        % Option-}
\catcode214=13 \def Ö{\sla}                      % Option-/; Option-V :
\catcode215=13 \def ×{\relax\ifmmode\check\else\expandafter\v\fi}
\catcode216=13 \def Ø{\"y}                     % Option-u y
\catcode217=13 \def Ù{\"Y}  		    % Option-u Y
\catcode218=13 \def Ú{\Leftarrow}       % Option-!
\catcode219=13 \def Û{\Leftrightarrow}       % Option-@ ; Option-# :
\catcode220=13 \def Ü{\relax\ifmmode\Rightarrow\else\sect\fi}
\catcode221=13 \def Ý{\sum}                  % Option-$
\catcode222=13 \def Þ{\prod}                 % Option-%
\catcode223=13 \def ß{\widehat}              % Option-^
\catcode224=13 \def à{\pm}                     % Option-&
\catcode225=13 \def á{\nabla}                % Option-(
\catcode226=13 \def â{\quad}                 % Option-)
\catcode227=13 \def ã{\in}               	% Option-W
\catcode228=13 \def ä{\star}      	      % Option-R
\catcode229=13 \def å{\sqrt}                   % Option-M
\catcode230=13 \def æ{\^E}			% Option-i E
\catcode231=13 \def ç{\Upsilon}              % Option-Y
\catcode232=13 \def è{\"E}    	   	 % Option-u E
\catcode233=13 \def é{\`E}               	  % Option-`  E
\catcode234=13 \def ê{\Sigma}                % Option-S
\catcode235=13 \def ë{\Delta}                 % Option-D
\catcode236=13 \def ì{\Phi}                     % Option-F
\catcode237=13 \def í{\`I}        		   % Option-`  I
\catcode238=13 \def î{\iota}        	     % Option-H
\catcode239=13 \def ï{\Psi}                     % Option-J
\catcode240=13 \def ð{\times}                  % Option-K
\catcode241=13 \def ñ{\Lambda}             % Option-L
\catcode242=13 \def ò{\cdots}                % Option-:
\catcode243=13 \def ó{\^U}			% Option-i U
\catcode244=13 \def ô{\`U}    	              % Option-`  U
\catcode245=13 \def õ{\bo}                       % Option-B ; Option-I :
\catcode246=13 \def ö{\relax\ifmmode\hat\else\expandafter\^\fi}
\catcode247=13 \def÷{\relax\ifmmode\tilde\else\expandafter\~\fi}
\catcode248=13 \def ø{\ll}                         % Option-< ; ^ Option-N
\catcode249=13 \def ù{\gg}                       % Option-> 
\catcode250=13 \def ú{\eta}                      % Option-h 
\catcode251=13 \def û{\kappa}                  % Option-k 
\catcode252=13 \def ü{\half}     		 % Option-Z 
\catcode253=13 \def ý{\Gamma} 		% Option-G 
\catcode254=13 \def þ{\Xi}   			% Option-X ; Option-T : 
\catcode255=13 \def ÿ{\relax\ifmmode{}^{\dagger}{}\else\dag\fi}

% hat, check, tilde, and bar have been defined to work in text as well.

\def\ital#1Õ{{\it#1\/}}	     % for italics in text: see Ó above
\def\un#1{\relax\ifmmode\underline#1\else $\underline{\hbox{#1}}$
	\relax\fi}

	% for unitalicized
	% accent under
\def\tdt#1{\on{\hbox{\bf .\kern-1pt.\kern-1pt.}}#1}   % À À À
\def\({\eqno(}

\def\refs{\sect{REFERENCES}\par\medskip \frenchspacing 
	\parskip=0pt \refrm \baselineskip=1.23em plus 1pt
	\def\ital##1Õ{{\refit##1\/}}}

% Young tableaux:  \upõ<a>{\õ<a>...\õ<b>}

\def\õ#1{
	\screwcount\num
	\num=1
	\screwdimen\downsy
	\downsy=-1.5ex
	\mkern-3.5mu
	õ
	\loop
	\ifnum\num<#1
	\llap{\raise\num\downsy\hbox{$õ$}}
	\advance\num by1
	\repeat}
\def\upõ#1#2{\screwcount\numup
	\numup=#1
	\advance\numup by-1
	\screwdimen\upsy
	\upsy=.75ex
	\mkern3.5mu
	\raise\numup\upsy\hbox{$#2$}}

% Some stupid little things that have to go at the end:
% make |,<,> OK in text

\catcode`\|=\active \catcode`\<=\active \catcode`\>=\active 
\def|{\relax\ifmmode\delimiter"026A30C \else$\mathchar"026A$\fi}
\def<{\relax\ifmmode\mathchar"313C \else$\mathchar"313C$\fi}
\def>{\relax\ifmmode\mathchar"313E \else$\mathchar"313E$\fi}

% relate notation

\def\dot{\dt}

%%%%%%%%%%%%%%%%%%%%%%%%%%%%%%%%%%%%%%%%

\paper

SIMPLIFYING ALGEBRA\\ IN FEYNMAN GRAPHS\\
Part I: Spinors

\author{G. Chalmers and W. Siegel\footnote{${}^1$}{
 Internet addresses: chalmers and siegel@insti.physics.sunysb.edu.}}\ITP

97-45

August 5, 1997

We present a general formalism for simplifying manipulations of spin 
indices of massless and massive spinors and vectors in Feynman
diagrams.   The formalism is based on covariantly reducing the number
of field  components in the action in favor of chiral/self-dual fields.    In
this paper we concentrate on calculational simplifications involving 
fermions in gauge theories by eliminating half of the components of
Dirac  spinors.  Some results are: (1) We find reference momenta for
massless fermions analogous to those used  for external gauge bosons. 
(2) Many  of the known supersymmetry  identities (tree and one-loop)
are seen  in a simple manner from the graphs. (3) Manipulations with
external  line factors for massive fermions are unnecessary. (4) Some of
the  simplifications for nearly maximally helicity violating gluonic
amplitudes  are built into the Feynman rules.   

Ü1. Introduction  

In the past decade there have been many advances in the art of  doing
perturbative gauge theory calculations.  (In [1] reviews of  these
methods are presented for massless QCD.)  Color ordering,  new
Feynman rules derived from string theory,  and techniques based  on
unitarity and analyticity  requirements on S-matrix elements have  been
important developments [2,3].  Notable among the new techniques  was
the introduction of reference momenta for external polarization 
tensors.  However, the simplifications for doing these calculations 
when external fermions are present have not been as dramatic, and  no
analogous simplifications for massive particles have been presented.  
In this work we fill this gap by introducing reference momenta for
massless external fermion lines analogous to those used  for gluons,
and similarly covariantly reduce the components of massive  external
line factors for spinors and vectors.  Many of the simplifications  are
introduced directly into the action by covariantly reducing the 
components of the fields themselves.  This leads to further 
simplifications for propagators and vertices.      

Twistors [4],  also known in gauge theory calculations as  ``spinor
helicity" [5], are formulated by writing all massless (on-shell) momenta
in terms of  commuting spinors, 
$$ 
k^2=0  
 \Rightarrow k_{\alpha \dot\alpha}=\pm k_\alpha k_{\dot\alpha}  \ .   
\(1)
$$  
Or in matrix notation
$$   
\langle k\vert = k^\alpha \ ,â \vert k\rangle= k_\alpha \ ;  ââ
 [k\vert = k^{\dot\alpha} \ , â\vert k] = k_{\dot\alpha} 
\(2)
$$ 
we write 
$$  
k=\pm \vert k\rangle [k\vert \ . 
\(3)
$$
Furthermore, polarization  vectors $\epsilon^{\pm}_{\alpha 
{\dot\alpha}}$ satisfying the required normalization conditions may 
also be represented in terms of commuting spinors as 
$$  
\epsilon^+_{\alpha\dot\alpha}(k) = -i{k_{\dot\alpha} p_\alpha \over 
 k^\beta p_\beta} 
\qquad\qquad 
\epsilon^-_{\alpha\dot\alpha}(k) = -i{k_\alpha p_{\dot\alpha}\over 
 k^{\dot\beta} p_{\dot\beta} }
 \ ,   
\(4) 
$$    
or in matrix notation, 
$$   
\epsilon^+ = -i {\vert p\rangle [k\vert \over \langle kp\rangle} 
\qquad\qquad  
\epsilon^- = -i {\vert k\rangle [p\vert \over [ kp]}  \ .  
\(5)
$$
Our conventions are such that $(\psi^\alpha)^\star=\psi^{\dot\alpha}$ 
and $(\psi_{\dot\alpha})^\star=-\psi_\alpha$.  
Because of gauge invariance the polarization spinor $p_\alpha$  is arbitrary
(but  $k^\alpha p_\alpha\neq 0$):
$$ ¶A_{ŒÀŒ} = »_{ŒÀŒ}ÂâÜ⶷_{ŒÀŒ}(k) = ik_{ŒÀŒ}Â(k) 
 âÜâ¶p_Œ = ½(k)k_Œ  + \xi(k) p_\alpha  \ ,
\(6) 
$$  
where $½(k)$ comes from the gauge transformation parameter 
$\lambda(k)$ while the scale parameter $\xi(k)$ comes from 
the normalization of the polarization vectors.   
 The perturbative calculations  can 
then be simplified by identifying the null vector 
 $p_{\alpha\dot\alpha}  = p_\alpha p_{\dot\alpha}$ with any other
massless (``reference") momentum  in the amplitude.   

This approach has been applied both to eliminate much algebra in 
intermediate stages of calculations and to reduce the expressions for the
final result of the amplitude.  In  some cases the resulting amplitudes are 
extremely simple, such as the case of pure gluon amplitudes where all 
external legs except one or two are similarly polarized  [6].  The simplicity 
of these expressions are now understood to arise from the relation of these
particular amplitudes to self-dual field theories [7].  The simplifications 
we present in this paper are based on actions that take advantage of 
chirality and self-duality.     

In our formulation reference momenta may 
also be given to external {\it fermion} lines.  This is interesting in that, 
unlike external vector bosons, there is no apparent local symmetry 
incorporating the fermion field.   We first integrate out half of the 
massive spin 1/2 fields coupled to  gauge fields:  In such theories with 
a labelling of
the fields according  to an SL(2,C) (van der Waerden)  Weyl-spinor 
notation, the massive spinor fields with dotted indices may be 
eliminated in favor of those with undotted ones; we treat the 
former fields as Lagrange multipliers.  Taking the massless limit 
of the resulting Lagrangian defines our massless theory;  the 
definition of the in and out states then allows us to present the 
reference momenta for the external fermion lines. 

In Section 2 we present the reformulation of the actions for massive 
and massless spin 1/2 fields.  We give the improved Feynman rules in 
section 3, including the reference momenta for external fermions.  In
section 4 we reproduce a well-known result:  We illustrate  our
technique with a sample calculation of the high-energy limit 
$e^+ e^- \rightarrow \gamma \gamma$ tree-level scattering process. 
The known supersymmetry  identities and their implication for the
vanishing of certain tree and  one-loop amplitudes is immediately
obvious from the action and  is discussed in Section 5.   Section 6
contains our final comments.  In the sequel we will apply similar
methods to vectors (abelian and nonabelian).

Ü2.  Actions  

The action for a massive particle in an external massless vector field
(Yang-Mills or electromagnetism) can be written as
$$  {\cal L} = ÐÆ^{ÀŒ}iá_{ŒÀŒ}Æ^Œ + 
   {m\over 2} (Æ^Œ Æ_Œ +{\bar\psi}^{ÀŒ} {\bar\psi}_{ÀŒ}) \ , 
\(7)$$  
where $á_{ŒÀŒ}=\partial_{ŒÀŒ}+iA_{ŒÀŒ}$ is our convention for the 
covariant derivative.  After 
eliminating the dotted field as auxiliary using the field equations 
($i\nabla_{\alpha{\dot\alpha}}  
 \psi^\alpha = -m{\bar\psi}_{ÀŒ}$)
we have [8] 
$$ 
{\cal L} = {1\over 2m} 
  \bigl( \nabla_\alpha{}^{\dot\alpha} \psi^\alpha \bigr) 
  \bigl( \nabla_{\beta\dot\alpha} \psi^\beta \bigr) +  
  {m\over 2}  
\psi^\alpha \psi_\alpha \ , 
\(8) 
$$ 
or
$$ 
{\cal L} = -{1\over 2m} Æ^Œ (õ +{i\over 2} F_º{}^© S_©{}^º -m^2) Æ_Œ =
	-{1\over 2m} Æ^Œ (õ -m^2) Æ_Œ -{i\over 2m} Æ^Œ F_Œ{}^º Æ_º \ ,  
\(9)$$   
for real representations of the gauge group.  Note that 
we have used the conventions where $V\cdot W={1\over 2} 
V^{\alpha\dot\alpha} W_{\alpha\dot\alpha}$ and $F_{\alpha\beta} 
= {1\over 2} \partial_{(\alpha}{}^{\dot\alpha}
A_{\beta)\dot\alpha}$.  When the real representation is complex plus
complex conjugate (e.g., QED or QCD),  we can write
$\psi^\alpha=(\chi^\alpha, \xi^\alpha)$ where 
$\chi$ and $\xi$ are in complex conjugate representations:   
$$
{\cal L} = -{1\over m} ^Œ (õ +{i\over 2}F_º{}^© S_©{}^º -m^2) \xi_Œ =
	-{1\over m} ^Œ (õ -m^2) \xi_Œ -{i\over m} ^Œ F_Œ{}^º \xi_º \ ,  
\(10)$$   
In addition, the same procedure can be applied for complex 
representations, i.e., parity violation, but the result is more 
complicated. 
Because the original action possessed a quadratic term $ÐÆ^2$, 
the elimination of $ÐÆ^{ÀŒ}$ produced only a trivial functional 
determinant factor in the path integral.  The overall 
$1/m$ can also be removed from the kinetic term by scaling the 
fermionic field 
$$ Æ £ åmÆ\ . 
\(11)$$  
It is interesting that upon setting the anti-self-dual part 
$F_{\alpha\beta}$ of the field strength to zero the action 
(9) loses its spin dependence, and the couplings of the 
fermions to the gauge fields are exactly what 
one obtains for scalars.  For this reason
there is much improvement in using the action (9) to compute S-matrix
elements involving external fermions; this will be demonstrated in later
sections.    

Our method is simple to implement in actual calculations; however, it  is not
equivalent to other known tricks that simplify $©$-matrix algebra.   Our 
reformulation of the original vertex is analogous to applying the Gordon
identity, 
$$  
{\bar u}(p) \gamma^a u(q)  
 = {1\over 2m} {\bar u}
\Bigl[ (p+q)^a
 + 2S^{ab} (q-p)_b \Bigr] u(q)   \ . 
\(12) 
$$ 
The rescaling of the fermion by the mass factor $\sqrt{m}$  occurs
naturally in the above when taking the massless limit.  The
``squared-propagator" trick, originally introduced by Feynman and
Schwinger, uses the fact that
$(Öá-m)(Öá+m)=õ+...$ is more analogous to the scalar theory.  Although
these manipulations give a convenient  rearrangement of $©$-matrices
along fermion lines, the number of $©$-matrices involved is the same.  

The idea of using fields with only undotted spinor indices has been used
previously for spin 1/2 fields in [8],  but advantage was not taken of the
simplification of external-line factors;  the fermion reference
momentum was not introduced.  More importantly, the $©$ matrices
were replaced by the 4-vector $§$-matrices;  in [8] the $§$-matrices
and their transposes were effectively treated as independent.  The
result of these manipulations does not change  the amount of
$©$-matrix algebra in actual calculations.

One of the major advantages of our method is that the $§$ matrices are
treated instead as self-dual tensors.   With the squared-propagator
trick, only even numbers of $©$-matrices appear, so the matrix algebra
consists of only 8 of the usual 16, namely $I$, $©_5$, and the spin
operators
$S_{ab}¾©^2_{[ab]}$.  When we restrict to chiral spinors, a further
reduction to just 4 matrices ($§$ and $I$) is achieved, since then $©_5$
is identified with the identity, and only the self-dual part of the spin
survives.   In the notation where representations are written as
$(m,n)$, where $m$ and $n$ are either integral or half-integral
(corresponding to $2m$ undotted indices and $2n$ dotted indices), these
$§$-matrices appear as $(1,0)$, not as
$(ü,ü)$.  Thus, the $§$-matrices we use are treated as spin, as in
nonrelativistic quantum mechanics, and not as part of $©_a$.  (I.e.,  they
are treated as bosonic, not fermionic.)   The action, however, is still
manifestly  Lorentz covariant:  SL(2,C), when restricted to just
undotted  indices, is indistinguishable from SU(2).  More generally, our
massive fields are all
$(m,0)$ representations.  We emphasize that the  resulting matrices
carry {\it only} an effective SU(2) algebra, and not  the usual
SU(2)$°$SU(2) associated with the covering group SL(2,C), since the
fields carry only undotted indices.   Also, instead of using explicit  Pauli
$§$-matrices (which are the Clebsch-Gordan coefficients for
$ü°ü=1¢0$), we use SU(2) spinor notation, which is simpler even in
nonrelativistic quantum mechanics.

Thus, where a fermion line in the old formalism gives a string of the form
$$ Ðu_f ÖV_1 ÖP_1 ÖV_2 ... ÖP_{n-1} ÖV_n u_i 
\(13) $$
for $n$ vertices $ÖV_j$ and $n-1$ propagators $ÖP_k$; the new way gives
$$ V_1 V_2 ... V_n \ .
\(14) $$
Matrix multiplication is now trivial, of the form
$$ AB = (AÉB)I +AðB \ ,
\(15) $$
rather than
$$ ©^{(m)}©^{(n)} = ©^{(m+n)} + ©^{(m+n-1)} + ... +I 
\(16) $$
for the antisymmetrized product $©^{(n)}$ of $n$ $©$ matrices, since any
2$ð$2 matrix contains only the singlet or (1,0) representations.  Similarly, for
two fermion lines the old method required Fierz identities, which are not
needed in our formalism.

There is also a major simplification for the case in which the
massless Yang-Mills fields that couple to the fermion field are (almost)
all of the same helicity:  Only the self-dual part $S_{\alpha\beta}$ 
of the spin operator appears in the action (9) in the magnetic-moment
coupling; the coupling of anti-self-dual Yang-Mills fields, which 
generate opposite helicity states, is spin-independent.   In calculating
amplitudes closer to self-dual  ones (i.e., MHV) the fermions resemble 
scalars in actual calculations.   

Because our method treats dotted and 
undotted indices in an asymmetric fashion, the action (9) is 
complex, and unitarity is not readily apparent.  However, because  
unitarity is present within the original action (7), it must persist after
integrating out the auxiliary  fermionic fields
${\bar\psi}^{\dot\alpha}$.  (Even in non-abelian gauge theories 
used with a complex gauge fixing term, for example
Gervais-Neveu gauge $L_{gf}={1\over\lambda}{\rm Tr}\bigl( 
\partial\cdot A + iA^2\bigr)^2$, unitarity is not immediately
obvious from the reality properties of the action.) 
 
For example, because only the 
self-dual part of the classical magnetic moment coupling appears 
in the action (9), it is not immediately obvious the complete 
contribution arises.  The effect of
this coupling may be found from the covariantized Pauli-Lubanski vector
$W_{\alpha\dot\beta} =  S_{\alpha\beta} \nabla^\beta{}_{\dot\beta}$. 
Commuting this operator  with the ``Hamiltonian" 
$$
H= õ  - {i\over 2}  F^{\alpha\beta} S_{\beta\alpha}   
\(17)
$$ 
gives the usual precessions 
$$  
-i{d\over d }W_{ŒÀº} = [ H, W_{\alpha\dot\beta} ] =
i W_\alpha{}^{\dot\delta} F_{{\dot\delta}{\dot\beta}}  
 + i W^\rho{}_{\dot\beta} F_{\rho\alpha} \ .  
\(18) 
$$  
as follows from truncation of the usual expressions that include 
$S_{{\dot\alpha}{\dot\beta}}$ terms for $W$ and $H$.  
We see that the precession of the 
spin as described by the covariantized Pauli-Lubanski vector
has contributions from both self-dual and anti-self-dual fields.  

We also note that the same method used to obtain (9) has also been applied
to the classical  mechanics and classical field theory of the massive
superparticle  (describing spins 0 and 1/2), and the subcritical (Liouville) 
superstring [9].  There the undotted indices are carried by the 
anticommuting coordinates.  Such ``chiral superspaces" are natural 
for self-dual supersymmetric theories of any spins [10].  

Ü3.  Feynman rules

The rules for the propagators and vertices can be read from 
the action (9) as usual.  The vector propagator is the usual one, the 
same as the scalar propagator up to index factors:
$$  
ÒA_{ºÀ·}(k)A_{©À½}(-k)Ô= {1\over k^2} C_{º©} C_{À·À½} \ .   
\(19)  
$$    
The fermion propagator is now also like a scalar propagator:
$$  
\langle \psi^\alpha(k)
\psi^\beta(-k) \rangle =  {1\over k^2+m^2} C^{\alpha\beta}  \ .    
\(20)
$$ 
Correlations between external $F$ type couplings are  
simple; for example, we have  
$$ 
ÒF_{Œº}(k)F_{©¶}(-k)Ô =  
 \f14 k_{(Œ}{}^{À·}ÒA_{º)À·}(k)A_{À½(©}(-k)Ôk_{¶)}{}^{À½}
	= \f14 C_{©(Œ}C_{º)¶} \ .  
\(21) 
$$  

The coupling of a gauge field to the charged spinors may be 
found from the Lagrangian (9) and appears in three types.  The 
fermion line either emits a gauge boson in a spin-independent 
fashion through the expansion of the gauge covariant box, or 
through the self-dual field strength.  In the former case we 
have the $e^+ e^- \gamma$ coupling 
$$ 
V_{\alpha{\dot\alpha}}^{\mu\nu} =ü
  e(k_1 - k_2)_{\alpha{\dot\alpha}} C^{\mu\nu}  
\(22a) 
$$ 
for a gauge vector $A_{\alpha{\dot\alpha}}$ and fermion indices 
$\mu$ and $\nu$.  The $e^+ e^- \gamma$ vertex coming from 
the $\psi^\alpha F_{\alpha\beta} \psi^\beta$ term in the Lagrangian 
is 
$$  
V_{\alpha{\dot\alpha}}^{\mu\nu} =  
   üe(k_1 - k_2)^{(\mu}{}_{\dot\alpha} ¶^{\nu)}_\delta \ .  
\(22b) 
$$ 
The external vector emitted through the 
chiral $F_{μ}(k)$-type coupling is a ``$-$" helicity state.  
The final vertex is the four-point $e^+ e^- \gamma\gamma$ 
coupling, 
$$  
V_{\alpha{\dot\alpha},\beta{\dot\beta}}^{\mu\nu} =  
 e^2 C_{\alpha\beta} C_{{\dot\alpha}{\dot\beta}} C^{\mu\nu} \ .  
\(23) 
$$ 
The couplings (22a) and (23) are of the same form as the 
scalar ones, but with additional indices $(\mu\nu)$ of the 
fermions contracted with the external fermion lines. 

It might appear that the Feynman rules for vertices are more complicated
than in the usual formalism, because there are more.  However, all
such terms arise in either method: here, directly in the action; by the usual
method, after performing Dirac algebra.  The advantage in the 
formulation presented here is that
the $\gamma$-matrix algebra has been performed once and for all in the
action itself, whereas in the usual method one must reshuffle the 
terms individually in each Feynman diagram at each vertex and propagator. 
Also, our form allows for more convenient comparison to other spins (e.g., for
supersymmetry).

In deriving the S-matrix elements in terms of the reduced 
action (9) we need to specify the in- and  outgoing line factors 
for the external fermions.  In the massive case external line 
factors are arbitrary; their choice corresponds to one of spin 
axis.  Unlike Dirac spinors, in our case massive external line 
factors are not needed to determine the four components in 
terms of two polarizations, since we have already reduced 
explicitly to only two.  With the usual methods, external 
line factors are needed to reduce the $4^n$ matrix elements 
for $n$ external fermion lines to $2^n$ appropriate to the 2 
polarizations of spin 1/2.  The alternative is to square the 
amplitude before performing Dirac algebra.  However, this 
can be cumbersome, especially when numerical evaluation of momentum
integrals is involved, since $N$ diagrams produce $N^2$ terms in the cross
section.  

In the  massless case, the line factors are determined by helicity:  The 
two helicity states for $\psi_\alpha$ are, with the
normalization $\epsilon^{+,\alpha} \epsilon^-_\alpha = 1$,
$$   
\epsilon_\alpha^+ = p_\alpha \qquad\qquad 
 \epsilon_\alpha^- = {q_\alpha \over p^\beta q_\beta} \ ,  
\(24)   
$$   
or in matrix notation
$$ ·^+ = |pÔââ·^- = {|qÔ\over ÒpqÔ} \ ,  
\(25) $$
in terms of an arbitrary twistor $q_\alpha$.  These states correspond 
to solutions to the field equations in the
original action (7), $\psi_\alpha=\epsilon_\alpha^+$  and
${\bar\psi}_{\dot\alpha}=p_{\dot\alpha}=  
-i\partial^\alpha{}_{\dot\alpha}\epsilon^-_\alpha$.  The 
relation to (9) can be seen, e.g., by introducing a $4$-component 
background spinor for (7) before integrating out the barred 
fields.  The 
ambiguity in choosing $q_\alpha$ is analogous to  choosing reference
momenta for gauge field polarization  vectors in the spinor helicity 
formalism (an independent derivation of a fermionic reference 
momenta was given in (11) ).  Note that the vector
polarizations are products of the spinor polarizations and their complex conjugates
(including normalization).
 In the remaining part of the section we illustrate the
simplifications obtained by using  these Feynman rules.  
  
The rules are simplest when the amplitudes possess external 
vector lines mostly of the same helicity (i.e., of the ``$+$" type).  If an
external gauge boson of helicity ``$-$" is emitted through an
$F_{\alpha\beta}$ coupling, we  can immediately apply twistor 
techniques to write the corresponding external-line factor with 
momentum $k^2=0$ in momentum space 
$$  
F_{Œº}(k) ={i\over 2} k_{(Œ}{}^{ÀŒ} A_{º) ÀŒ}(k) = k_Œ k_º \ .  
\(26) 
$$  
or
$$ F = |kÔÒk|  
\(27) $$
Because of gauge  invariance it is independent of the reference
momentum chosen for the  vector boson. 

As an example of the reduced spinor algebra associated with  these
couplings, we consider the case of a single fermion  line with a number
of attached external gauge bosons.  All the algebra associated with the
indices of the external fermions comes from the $F$ coupling.  We 
immediately obtain  with the propagator (20) the contracted vertex
algebra  
$$ 
F_\alpha{}^º(k_1) F_º{}^\sigma (k_2) F_\sigma{}^\rho (k_3)  
  \ldots =
\vert 1\rangle Ò12ÔÒ23Ô...,ââ\langle i\vert =k_i{}^\alpha, \quad 
 \vert j\rangle = k_{j\alpha} 
\(28) 
$$ 
from the couplings.   The numbers 1,2,... label the consecutive $F$'s that
appear in the  product of external $F$ fields attached to the fermion
line; the  remaining spin independent couplings appear as scalars and 
do not effect the matrix algebra associated with the SU(2)  indices.  If
the fermion line is a closed loop then the $n$ $F$'s  contract to a cyclic
product 
$$ 
P=Ò12ÔÒ23Ô...Òn1Ô \ ;  
\(29)
$$
if the line is open, then the product is terminated on either end with 
the polarization spinors $Òf|$ and $|iÔ$ of the corresponding 
external fermion fields, 
$$
P=Òf1Ô...ÒniÔ 
\(30) $$   
with the helicity states (24) in the massless case.    

Lastly, we comment on the use of the Lagrangian (9) for  fermions in
massive QED calculations.   Consider for example the scattering of a
fermion/anti-fermion pair going into $n-2$ vectors of the same 
helicity.  Because the $F_\alpha{}^\beta$ vertex does not enter into the
calculations  (it generates the ``$-$" helicity states) we find   that the
expressions for the graphs are the same as the ones  obtained for
$n-2$  photons emitted along a scalar line.  The Feynman rules in this
particular example completely avoid the usual Dirac  matrix algebra,
external line factors, and  field equations usually used to simplify the
algebra.  Similar simplifications occur for amplitudes with a general  set
of polarizations for the external vectors, or virtual photons  connecting
to other fermion lines.      

Ü4. Massless Example 
 
In this section we reproduce two very simple QED (and QCD) scattering 
processes using these rules (in the massless or high-energy limit).  

A classic scattering example is $e^+ e^- \rightarrow \gamma^- 
\gamma^+$.   
A similar analysis will provide the massless QCD amplitude 
for $q{\bar q}\rightarrow g^+ g^-$ after including the color factors 
and the additional four-point contact term within the non-abelian 
$F_\alpha{}^\beta$ spin-dependent interaction.  
We label the incoming fermion line with momentum $k_1$ and 
the outgoing one with $k_2$; the photons with outgoing 
momentum $k_3$ and $k_4$ have helicity $-$ and $+$ respectively.  
In a color-ordered format, there are three diagrams --- one with an 
ordered labelling 1234 of the external legs (containing an $s_{14}$-channel), 
one with the crossed legs 1243 (containing a $s_{24}$-channel), and an 
additional one from the $e^+ e^- \gamma\gamma$  four-point 
contact term within the covariant box.  
 
Because the outgoing photons have opposite helicities, there 
is at most one emission of a vector through the spin-dependent 
$F$-type coupling (which generates a minus helicity vector state).  
The first ordered diagram receives contributions from the 
emission of a ``$-$" state through either a 
$\psi^\alpha F_\alpha{}^\beta\psi_\beta$ coupling or the 
$A\cdot \partial$ 
within the covariantized box; we label the contributions as $T_0$ and 
$T_1$ with the index labelling the number of $F$ couplings.  The 
Feynman rules give the expressions for the diagrams:  
$$
T^{(1234)}_0 = e^2 {\langle 2q \rangle \langle 32\rangle [2k]  
 \langle 1p\rangle 
 \over \langle q1\rangle \langle 14\rangle [k3] \langle p4\rangle} 
\qquad 
T^{(1243)}_0 = e^2 {\langle 2q \rangle \langle 42\rangle [2p]  
 \langle 1k\rangle 
 \over \langle q1\rangle \langle 13\rangle [p4] \langle k3\rangle} 
\(31)
$$ 
and
$$  
T^{(1234)}_1 = e^2 {\langle 23\rangle \langle 3q\rangle  
 \langle 1p\rangle \over 
 \langle q1\rangle \langle 14\rangle \langle p4\rangle} 
\qquad 
T^{(1243)}_1 = -e^2  {\langle 23\rangle \langle 3q\rangle  
 \langle 2p\rangle \over 
 \langle q1\rangle \langle 24\rangle \langle p4\rangle} \ ,
\(32)
$$  
where $k$ and $p$ are the reference momenta for the photons with 
momentum $k_3$ and $k_4$, respectively, and $q$ is that for the fermion
with momentum $k_1$.  The  next contribution arises from the four-point
contact term and is 
$$  
C = e^2 {\langle 2q\rangle\over \langle q1\rangle} 
{\langle 3p\rangle [4k]\over [3k] \langle 4p\rangle} \ .  
\(33)
$$
The final result after adding up all the contributions is 
guaranteed to be independent of the reference momentum.  However, 
we should choose them to simplify the intermediate steps in 
the calculation.  

For example, upon taking $q=k_2$ we eliminate $T^{(1234)}_0$, 
$T^{(1243)}_0$, and $C$.  Choosing the reference momentum $p=k_2$ 
for the ``$-$" helicity outgoing photon eliminates $T^{(1243)}_1$.  The 
entire result for the amplitude then arises from the $T^{(1234)}_1$ 
contribution; it is 
$$  
A(e^+,e^-;k_3^-,k_4^+) 
 = e^2 {\langle 23\rangle^2\over \langle 14\rangle  \langle 24\rangle} 
 = - e^2 {[14] \langle 23\rangle \over \langle 42\rangle [23]}  
\ ,  \(34) 
$$  
where we have multiplied by $[23]/[23]$ and used $k_2Ék_3=k_1Ék_4$ to
show agreement with the result obtained by conventional techniques. 
In practice, the reference momenta are chosen at the beginning of the
calculation, so only the single graph $T_1^{(1234)}$ is actually
calculated.  Although the amplitude here is a relatively easy one to
evaluate with  the more conventional techniques, our evaluation did not
involve any gamma matrix algebra; for higher-point diagrams this is a
significant advantage.  Furthermore, we expect that  in one-loop
amplitudes the fermionic reference momenta will  lead to significant
calculational advantages.     

Ü5. Supersymmetry Identities  

In this section we rederive several  of the known supersymmetry
identities relevant to maximally  helicity violating amplitudes [12,1] and
their relation to self-dual  Yang-Mills theory in $2+2$ dimensions.  The
reformulation of the  gauge theory we present is naturally suited to
deriving these  identities.  

First, consider the scattering at tree-level of a $q {\bar q}$  pair into a
series of 
$+$ helicity outgoing non-abelian vectors, $A(q, {\bar q},g^+,\ldots, 
g^+)$.  This example is particularly simple to describe  with our rules
because an off-shell field
$\langle A_{\alpha\dot\alpha}\rangle$ evaluated at tree-level  between
on-shell self-dual states is itself self-dual [13]; i.e., any tree amplitude
with a number of external legs of the same helicity  attached to the
fermion line vanish if it is coupled through an 
$F_{μ}=0$ term in the Lagrangian (9).  Along the fermion line the
vectors are then emitted through the (scalar-type) coupling found from
expanding the covariantized box 
$\psi^\alpha õ \psi_\alpha$.  Because the gluons are emitted through  a
scalar-type coupling the incoming/outgoing fermions contract 
immediately to give $\langle fi\rangle= \langle 1q\rangle /  
\langle 2q\rangle$.  (The $q$,${\bar q}$ lines possess momenta 
$k_1$, $k_2$.)  The entire contribution is immediately seen to vanish
upon taking $q=k_1$.  A  similar analysis shows that amplitudes with
any number of fermions and vectors containing a maximally helicity
violating assignment of  polarizations vanish.    
   
Next, the well-known relation for the partial amplitude, 
where $\phi$ and $\bar\phi$ are complex scalars,  
$$  
A_{\rm tree}(e^+,1^+,\ldots,j^-,\ldots,(n-1)^+,e^-) 
=  {\langle j1\rangle \over \langle jn\rangle } 
A_{\rm tree}({\bar\phi}_1, 1^+,\ldots,  
  j^-,\ldots,  (n-1)^+,\phi_n)  
\(35) 
$$  
may also be easily found with the use of our new Feynman rules.  
In this example, for simplicity we consider abelian gauge 
fields only. 
The coupling of scalars to the gauge field is similar to that of the 
fermions; however, there is no spin dependent $F$ term: 
$$ 
{\cal L} = -{\bar\phi} (õ - m^2) \phi  \ ,  
\(36)  
$$   
In deriving the $e^+ e^-$ amplitude the ``$-$" helicity $j^{\rm th}$ 
vector along the fermion line may be emitted through a
$\psi^\alpha F_\alpha{}^\beta \psi^\beta$ vertex or through the
fermion-vector coupling  found  in the  expansion of the covariant box.   In
the former  case, within each diagram the external fermion line factors 
do not contract with any of the algebra associated with the vector
emission except for the single $F_{\alpha\beta}(k_j)$ field to 
give, as described in (28),  
$$  
{\langle q j\rangle \langle j n\rangle \over 
 \langle q 1\rangle }  \ .
\(37) 
$$  
With the simple choice of $q=k_j$, all of these diagrams are set 
to zero and do not contribute to the amplitude.   The external 
fermion line factors in the 
remaining $e^+ e^-\rightarrow$ vector diagrams, i.e., those without 
any explicit $F_{\alpha\beta}$ couplings, 
contract directly to give
$$  
{\langle q n\rangle \over \langle q 1\rangle}\Bigr|_{q=k_j} 
 = {\langle j n\rangle \over \langle j 1\rangle}   \ .
\(38) 
$$  
In comparing the diagrams contributing to the $e^+ e^- \gamma^+ 
\ldots \gamma^+$ and $\phi\phi\gamma^+\ldots\gamma^+$ 
amplitudes, we find the relation in (35).

The last identity on S-matrix elements we discuss is one which relates
different MHV amplitudes at one-loop.  These S-matrix elements  
are known to satisfy, for any number of external legs, the 
identity 
$$  
A_{n;1}^{[1]}=A_{n;1}^{[0]}=-A_{n;1}^{[1/2]}   
  (g^+,\ldots,g^+) \ . 
\(39)
$$ 
The index $[j]$ labels the spin content of the internal states 
within the loop (complex scalar, Weyl fermion, and gluon).  By 
the indices $n;1$ we mean the single-trace structure in a 
non-abelian gauge theory using the color-ordered Feynman 
rules [2] derived from the action (9).  The result (39) is normally 
found through a supersymmetric identity which states that the 
contribution of a virtual supersymmetric multiplet to the scattering 
gives zero, i.e., $A_{n;1}^{[N>0]}(g^+,\ldots,g^+)=0$; as discussed in [3] 
taking linear combinations of states at one-loop in different 
supersymmetric multiplets then gives (39).   

We examine the derivation of (39) in the following.  Clearly the only violation
of this identity between $j=1/2$ and $j=0$ contribution  ($j=1$ will be
discussed in the  sequel paper) can come from the coupling of trees from the
anti-self-dual field strength $F_{μ}$ to the loop.  However, as in  case of the
first identity discussed, we may take $F_{μ}=0$ in the amplitudes
$A_{n;1}^{[j]} (g^+,\ldots,g^+)$.  The reduced action formulation we present in
this work is thus naturally suited to describing self-dual Yang-Mills theory, 
which has an S-matrix coinciding with the Wick-rotated one-loop  MHV gauge
theory amplitudes. 
 
Ü6. Discussion 

For completeness we also give the reduced Lagrangian describing 
the theory of fermions obtaining their mass through a Higgs 
effect.  The Yukawa coupling of the Majorana fermion is,  
$$  
{\cal L}_{\rm y} = {1\over 2} \lambda ÄÆ^Œ Æ_Œ +  
   {1\over 2} \lambda ÐÄÐÆ^{ÀŒ} {\bar\psi}_{ÀŒ} \ . 
\(40) 
$$
In this theory, integrating out the dotted spinor components of the 
Lagrangian (7) with the mass terms replaced by (40) leads to
nonlinearities in the final action.   We find, after eliminating $ÐÆ$ and 
rescaling the $Æ$-field as in (11), the fermion contributions 
$$ {\cal L}=-{1\over 2} Æ^Œ(õ -\lambda^2 ÐÄÄ)Æ_Œ -{i\over 4} 
 Æ^Œ F_Œ{}^º Æ_º
$$ 
$$ +\f14 Æ^Œ Æ_Œ (á\ln ÐÄ)^2
	-üÆ^Œ Æ_Œ (õ\ln ÐÄ) -üÆ^Œ (á_{(Œ}{}^{À©}\ln ÐÄ)á_{º)À©}Æ^º  \ .
\(41) 
$$   
The expression (41) is defined by a perturbation about 
the vacuum value $Ò\phiÔ$ of the Higgs particle.  Similar 
simplifications for amplitude calculations, for example in the 
electroweak sector of the standard  model, are expected using 
the reduced action above.    

We expect that the use of these rules will aid substantially 
in the future computation of higher-point loop amplitudes 
involving external fermions.  The simplifications obtained 
in eliminating complicated intermediate algebra in gluon 
scattering amplitudes should persist in these cases as well.  
Furthermore, it would be interesting to generalize these 
examples of reduced Lagrangians to theories containing 
higher spin fields.  We have already looked at the case of 
spin 1 and found similar simplifications; details will be 
presented elsewhere.  As a final note, it would be interesting 
to find the local symmetry responsible for the ambiguity in 
choosing the fermionic reference momenta; the analogous 
invariance responsible for the simplifications involving the 
vector reference momenta is gauge symmetry.  The appearance 
of such fermionic gauge symmetry is suggested by a first-quantized 
approach to a spin-$1/2$ field [14].   

ÜACKNOWLEDGMENTS

This work was supported in part by the National Science Foundation
Grant No.¼PHY 9722101.  We thank Martin Ro×cek for useful discussions.  

\refs

£1 M. Mangano and S.J. Parke, ÓPhys. Rep.Õ É200 (1991) 301;\\
	Z. Bern, L. Dixon, D.A. Kosower, hep-ph/9602280,
	ÓAnn. Rev. Nucl. Part. Sci.Õ É46 (1996) 109.

£2 Z. Bern and D.A. Kosower, \PRD 38 (1988) 1888, \PR 66 (1991) 1669,
	\NP 379 (1992) 451, 562.

£3 Z. Bern, G. Chalmers, L. Dixon, and D.A. Kosower, hep-ph/9312333,
	\PR 72 (1994) 2134;\\
	Z. Bern, L. Dixon, D.C. Dunbar, and D.A. Kosower, hep-ph/9403226,
	\NP 425 (1994) 217, hep-ph/9409265, \NP 435 (1995) 59;\\
	Z. Bern and G. Chalmers, hep-ph/9503236, \NP 447 (1995) 465.

£4 R. Penrose, ÓJ. Math. Phys.Õ É8 (1967) 345, 
	ÓInt. J. Theor. Phys.Õ É1 (1968) 61;\\
	M.A.H. MacCallum and R. Penrose, ÓPhys. Rep.Õ É6C (1973) 241;\\
	A. Ferber, \NP 132 (1978) 55.

£5 P. De Causmaecker, R. Gastmans, W. Troost, and T.T. Wu, 
	\NP 206 (1982) 53;\\
	F.A. Berends, R. Kleiss, P. De Causmaecker, R. Gastmans, W. Troost,
	and T.T. Wu, \NP 206 (1982) 61;\\
	Z. Xu, D.-H. Zhang, and L. Chang, \NP 291 (1987) 392;\\
	J.F. Gunion and Z. Kunszt, \PL 161 (1985) 333;\\
	R. Kleiss and W.J. Sterling, \NP 262 (1985) 235. 

£6 S.J. Parke and T. Taylor, \NP 269 (1986) 410, \PR 56 (1986) 2459;\\
	F.A. Berends and W.T. Giele, \NP 306 (1988) 759;\\
	Z. Bern, G. Chalmers, L. Dixon, and D.A. Kosower, hep-ph/9312333, 
	\PR 72 (1994) 2134;\\
	G.D. Mahlon, hep-ph/9312276, \PRD 49 (1994) 4438.

£7 W.A. Bardeen, ÓProg. Theor. Phys. Suppl.Õ É123 (1996) 1;\\
	D. Cangemi, hep-th/9605208, \NP 484 (1997) 521;\\
       G. Chalmers and W. Siegel, hep-th/9606061, \PRD 54 (1996) 7628.

£8 L. M. Brown, ÓPhys. RevÕ É111 (1958) 957;\\
	M. Tonin, ÓNuo. Cim.Õ É14 (1959) 1108.

£9 W. Siegel, hep-th/9503173, \PRD 52 (1995) 3563.

£10 W. Siegel, hep-th/9205075, \PRD 46 (1992) R3235.

£11 A.G.\ Morgan, Phys.\ Lett.\ B351:249 (1995), hep-ph/9502230.

£12 M.T. Grisaru, H.N. Pendleton, and P. van Nieuwenhuizen, \PR 15
	(1977) 996; \\ 
	M.T. Grisaru and H.N. Pendleton, \NP 124 (1977) 333;\\
	M.T. Grisaru and W. Siegel, \PL 110B (1982) 49.

£13 M.J. Duff, C.J. Isham, \NP 162 (1980) 271.  

£14 N. Berkovits, M. T. Hatsuda, W. Siegel, \NP 371 (1992) 434.  

\bye